\documentclass[twocolumn,resetfootnote]{aastex7}

\usepackage[utf8]{inputenc}
\usepackage{tipa}
\usepackage{combelow}
\usepackage{savesym}
\savesymbol{tablenum}
\usepackage{siunitx}
\restoresymbol{SIX}{tablenum}
\usepackage{breqn}
\usepackage{tabularx}
\usepackage{placeins}

\begin{document}

\title{Prediction and observation of a stellar occultation by Haumea's satellite Namaka
}

\author[orcid=0000-0002-6085-3182,sname='Rommel']{Flavia L. Rommel}
\affiliation{Florida Space Institute, University of Central Florida, 12354 Research Parkway, Orlando, FL 32826, USA}
\email[show]{flavialuane.rommel@ucf.edu}

\author[orcid=0000-0002-1788-870X,sname='Proudfoot']{Benjamin C. N. Proudfoot}
\affiliation{Florida Space Institute, University of Central Florida, 12354 Research Parkway, Orlando, FL 32826, USA}
\email{}

\author[orcid=0000-0002-6117-0164]{Bryan J. Holler}
\affiliation{Space Telescope Science Institute, 3700 San Martin Drive, Baltimore, MD 21218, USA}
\email{}

\author[orcid=0000-0002-8690-2413]{Jose L. Ortiz}
\affiliation{Instituto de Astrof\'isica de Andaluc\'ia (CSIC), Glorieta de la Astronom\'ia s/n, 18008 Granada, Spain}
\email{}

\author[orcid=0000-0003-2132-7769]{Estela Fernández-Valenzuela}
\affiliation{Florida Space Institute, University of Central Florida, 12354 Research Parkway, Orlando, FL 32826, USA}
\email{estela@ucf.edu}

\begin{abstract}
Stellar occultations are an ideal way to characterize the physical and orbital properties of trans-Neptunian binary systems. In this research note, we detail the prediction and observation of a stellar occultation observed with NASA's IRTF on March 16$^{\mathrm{th}}$, 2025 (UT), with drop-outs from both the dwarf planet Haumea and its smaller satellite Namaka. This occultation places a lower limit of 83 $\pm$ 2 km on Namaka's diameter. We also discuss the possibility that this detection could help to constrain the orbit of Namaka, measure Haumea's gravitational harmonics, and provide a path to measuring the internal structure of Haumea. 
\end{abstract}


\section{Introduction} 
\label{sec:intro}

The study of trans-Neptunian binary and multiple systems has revealed a wealth of information about the processes of planetesimal formation, planetary migration, and collisional evolution in the outer solar system \citep[e.g.,][]{Noll2020}. Determining a satellite's orbit around its primary allows for the calculation of the system’s mass and, when combined with size estimates from radiometric techniques or stellar occultations, enables the derivation of system density—key for understanding composition and internal structure. Direct size measurements of satellites are valuable for computing visible geometric albedos, which enable a broad-strokes comparison of the surfaces of the system components. These quantities offer valuable constraints on the formation of the system, whether through gravitational collapse by the streaming instability, capture, giant impact, or rotational fission \citep[e.g.,][and references therein]{Noll2020,Bernstein2023}.

A reliable means of measuring the sizes of trans-Neptunian objects (TNOs) is the stellar occultation technique. A stellar occultation occurs when a solar system object passes in front of a background star, casting a shadow measurable with telescopes equipped with fast-readout cameras. The duration of the flux drop-out in an occultation light curve is converted to a projected distance on the sky, known as a ``chord,'' using the velocity of the event. To constrain the projected shape of the object, at least three chords are required, and combination with rotational light curve observations is essential for determining the rotation phase at the time of the occultation and the 3D shape. However, single-chord occultations are still valuable, as they provide limits on physical properties and additional astrometric measurements for the orbit solution, decreasing the uncertainty of the predicted path for future events.

In this research note, we present the results of a single-chord stellar occultation by Namaka, the smallest known satellite of the TNO dwarf planet (136108) Haumea, obtained with NASA's Infrared Telescope Facility (IRTF) in Hawaii. This is the first reported occultation chord obtained across Namaka, and thereby the first constraint on its size.

\begin{figure*}
    \centering
    \includegraphics[width=0.9\textwidth]{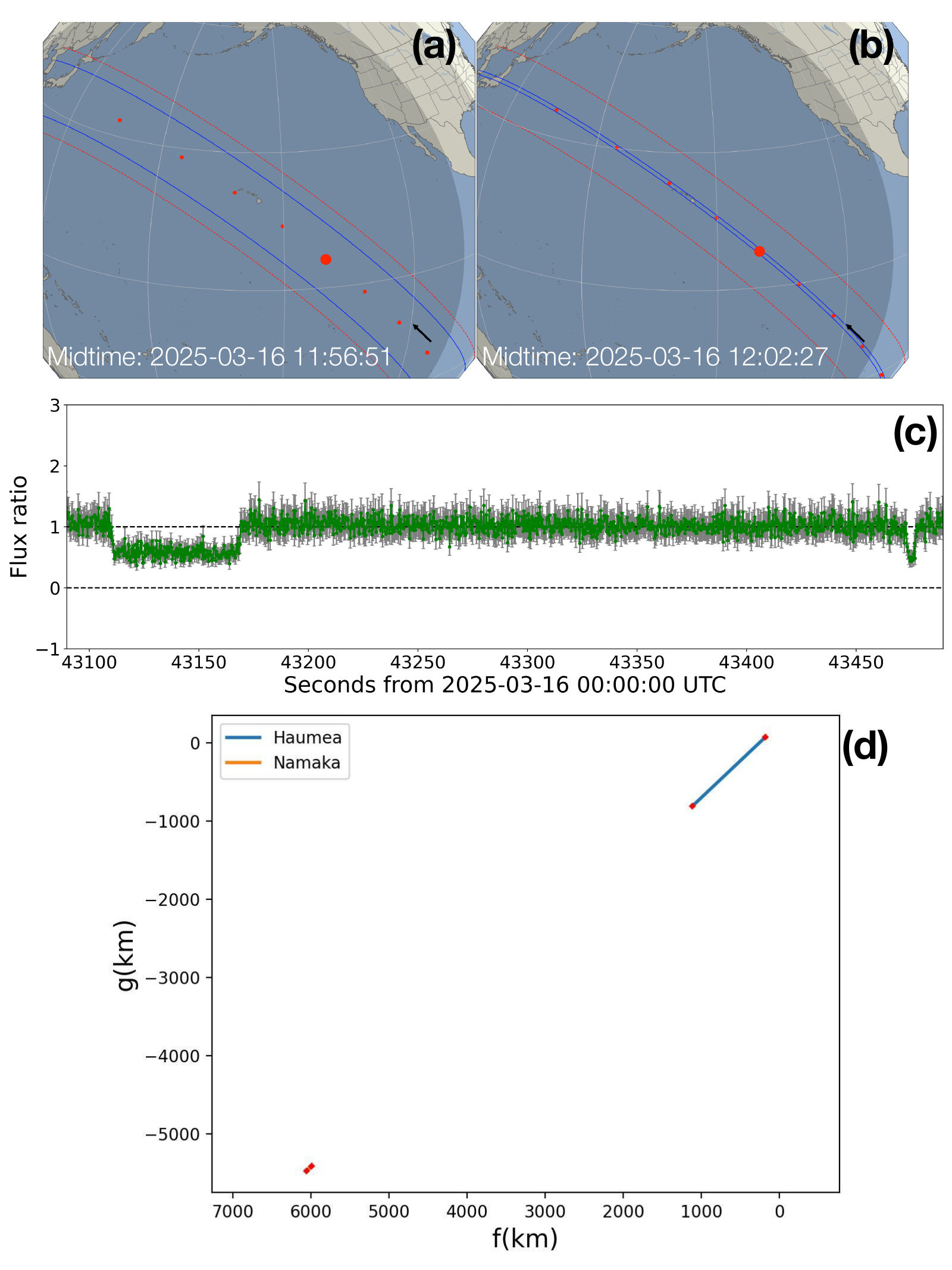}
    \caption{Predictions of the stellar occultations by (a) Haumea and (b) Namaka. Solid blue lines denote the estimated diameters of the objects and dashed red lines show the 1-$\sigma$ uncertainties. Panel (c) presents the normalized occultation light curve, and panel (d) presents the on-sky projection of the chords for Haumea and Namaka.}
    \label{fig}
\end{figure*}

\section{Prediction}
\label{sec:prediction}
We searched for potential events involving Namaka using the orbit solution from \cite{proudfoot2024beyond} together with the ephemeris of Haumea from NIMA\footnote{\url{https://lesia.obspm.fr/lucky-star/obj.php?p=1198}} \citep[][]{desmars2015orbit} and the \textsc{sora} tool \citep[][]{gomes2022sora}. We neglect the small barycentric wobble of Haumea, assuming that NIMA refers to Haumea's center. We predicted that Namaka would occult the same star as Haumea on March 16$^{\mathrm{th}}$, 2025 (UT). The occultation by Haumea was also predicted by the Lucky Star collaboration\footnote{\url{https://lesia.obspm.fr/lucky-star/occ.php?p=144652}}. Our predictions for the Haumea and Namaka occultations are shown in panels (a) and (b) of Figure \ref{fig}. 


\section{Observations and Data Analysis}

The event was recorded from the IRTF using the MORIS instrument \citep{Gulbis2011}, with an exposure time of 0.3 seconds and 2$\times$2 binning. The GPS pulse was set once per exposure to ensure that the image times were synchronized with UTC. Each FITS file comprised 200 frames, which we split using the {\tt astropy} v6.1.3 library. Relative aperture photometry was performed on each frame using \textsc{praia} \citep{Assafin2023} and featuring concentric apertures with radii of 3.0, 13, and 18 pixels. The inner aperture measured the stellar flux, and the sky annulus provided a median of sky brightness contributions.

After background subtraction, the flux of the target star was divided by the comparison star to reduce atmospheric transparency contributions. Given Haumea's brightness, a drop of $45 \%$ was expected; therefore, the light curve was normalized to unity outside the events and to 0.55 during the occultations. The normalized flux ratio as a function of time is shown in panel (c) of Figure \ref{fig}. The times of ingress and egress were derived using the $\chi^2$ test as implemented in the \textsc{sora} library. The best fit corresponded to $\chi^2_{min}$ and the 1-$\sigma$ uncertainties were calculated from $\chi^2_{min}+1$. The fit quality was assessed using $\chi^2_{pdf} = \chi^2_{min}/(N-M)$, where $N$ and $M$ were the number of measurements and free parameters, respectively. Our light curve fits provided $\chi^2_{pdf}=0.89$ and $\chi^2_{pdf}=0.97$ for the Haumea and Namaka events, respectively; $\chi^2_{pdf} = 1$ is the ideal result.

\section{Discussion and Conclusions}
\label{sec:discussion}

We detected the occultation of both Haumea and Namaka, resulting in chords of 1264 $\pm$ 2 km and 83 $\pm$ 2 km in length, respectively. Uncertainties were calculated at the 1-$\sigma$ level. The relative timing difference between the predicted and observed mid-times was only 0.7 $\pm$ 0.2 s, validating the Namaka orbit solution. With a single chord, we can only place a lower limit on Namaka's diameter of $D$ $\geq$ 83 $\pm$ 2 km. Assuming a spherical shape and considering a mass of  $1.18^{+0.25}_{-0.25} \times10^{18}$ kg \citep{proudfoot2019modeling}, we obtain a density of $\rho \leq 4000$ kg m$^{-3}$. For a more precise determination of Namaka's density, we will need future multi-chord occultations with the combination of its rotational light curve. Determining Namaka's density is pivotal to further understanding the formation of the only known collisional family in the trans-Neptunian region.

Combining our detection with the observations of Hi'iaka during two 2021 stellar occultations \citep[e.g.,][]{Fern2021}, it is now possible to provide precise updates to the Haumea system ephemeris. Notably, as was attempted in \citet{proudfoot2024beyond}, it may now be possible to independently measure both the masses of Hi'iaka and Namaka, as well as measure Haumea's gravitational harmonics. These gravitational harmonics---most crucially $J_2$---are related to Haumea's shape and internal structure. Since Haumea's shape was precisely measured by a stellar occultation \citep[][]{ortiz2017size}, the harmonics provide a probe into Haumea's interior. One geophysical model of Haumea's interior with a hydrated rocky core comprising 83\% of Haumea's volume predicts $J_2 \approx 0.16$ \citep[][]{dunham2019haumea}. If partially differentiated, Haumea could have a larger $J_2$, with an undifferentiated model predicting $J_2 \approx 0.24$. Thus, $J_2$ is a proxy for measuring the core size of Haumea. We leave a more detailed analysis to a follow-up paper.

\vspace{-5mm}


\begin{acknowledgments}

FLR and BCNP acknowledge the support of the UCF Preeminent Postdoctoral Program (P3). BJH acknowledges support from NASA SSO 19-SSO19\_2-0061. EFV acknowledges support from the Space Research Initiative. JLO acknowledges support from Spanish projects PID2020--112789GB-I00 (AEI and Proyecto de Excelencia de la Junta de Andalucía PY20-01309) and grant CEX2021-001131-S (MCIN/AEI/10.13039/501100011033). This work was based on data from the Infrared Telescope Facility, which is operated by the University of Hawaii under contract 80HQTR24DA010 with the National Aeronautics and Space Administration.

\end{acknowledgments}

\begin{contribution}

{{{EFV was the PI of the IRTF observing program (2025A043) through which this stellar occultation was observed. BCNP and JLO generated the predictions of the occultations. FLR and BJH carried out the remote observations. FLR performed the photometry to obtain the occultation light curve. FLR, BCNP, BJH, and EFV prepared the manuscript.}}}


\end{contribution}

\bibliography{all}{}
\bibliographystyle{aasjournal}



\end{document}